\documentclass[12pt]{article}
\usepackage{epsfig}
\usepackage{amssymb}
\usepackage{lscape}
\usepackage{cleveref}

\begin{document}

\title{Are the beginning and ending phases of epidemics provided by next
generation matrices? -- Revisiting drug sensitive and resistant tuberculosis
model\\
\bigskip }
\author{Hyun Mo Yang\thanks{%
Corresponding author: hyunyang@ime.unicamp.br; tel/fax: + 55 19 3521-6031}
\bigskip \bigskip \\
UNICAMP -- IMECC -- DMA \\
Pra\c{c}a S\'{e}rgio Buarque de Holanda, 651\\
CEP: 13083-859, Campinas, SP, Brazil}
\date{ }
\maketitle

\begin{abstract}
In epidemiological modelings, the spectral radius of the next generation
matrix evaluated at the trivial equilibrium was considered as the basic
reproduction number. Also, the global stability of the trivial equilibrium
point was determined by the left eigenvector associated to that next
generation matrix. More recently, the fraction of susceptible individuals
was also obtained from the next generation matrix. By revisiting drug
sensitive and resistant tuberculosis model, the gross reproduction number
and the fraction of susceptible individuals are calculated. Hence, the next
generation matrices shed light to the evolution of the dynamics: the
beginning of the epidemics via the reproduction number and the approaching
to the epidemics level via the asymptotic fraction of susceptible
individuals.

\bigskip

\end{abstract}

\section{Introduction}

In epidemiological modelings, in general, there is a unique threshold, which
is called the basic reproduction number (denoted by $R_{0}$). This number is
taken as the intensity at which epidemics spread out when one case is
introduced in a completely susceptible population. Additionally, by
evaluating the equilibrium value of the fraction of susceptible individuals
(denoted by $s^{\ast }$), this quantity could be expressed as $s^{\ast
}=1/R_{0}$. In another words, the inverse of the basic reproduction number
predicts the final size of an epidemics. Remembering that the basic
reproduction number is obtained from the stability analysis of the trivial
equilibrium point (denoted by $P^{0}$, which describes the absence of
epidemics), it is expected that both initial ($R_{0}$) and final ($s^{\ast }$%
) phases of epidemics can be obtained from this analysis, and, as a
consequence, it is not necessary the calculation of the fraction of
susceptible individuals $s^{\ast }$ in the steady state.

One of\ the approaches to determine the stability of the trivial equilibrium
point is evaluating the spectral radius of the corresponding next generation
matrix, which is taken as the basic reproduction number \cite{diekman}.
Instead of this spectral radius, the basic reproduction number is assumed to
be the sum of the coefficients of the characteristic equation corresponding
to the next generation matrix, which was proposed in \cite{yang1} and proved
in \cite{yang2}. In susceptible - exposed - infectious - recovered (SEIR)
model, there are two distinct characteristic equations corresponding to two
different next generation matrices, but the sum of the coefficients results
in the same threshold, which is linked to the basic reproduction number $%
R_{0}$ (hence, the second threshold $s^{\ast }$ appears implicitly, see
Discussion). But in SIR model, there is only one characteristic equation
resulting in a unique threshold, and the relationship $s^{\ast }=1/R_{0}$
comes out from the steady state value of the fraction of the susceptible
individuals. However, some models taking into account an additional route of
infection besides the infection of susceptible from an infectious individual
presents two different thresholds. In \cite{yang1} two procedures were
presented aiming the calculation of these two thresholds, which are the
gross reproduction number (denoted by $R_{g}$, with $R_{g}=R_{0}+R_{a}$,
where $R_{a}$ is an additional reproduction number) and the fraction of the
susceptible individuals $s^{\ast }$. Due to appearance of two different
thresholds, there is not the inverse relationship between gross reproduction
number and fraction of susceptible individuals, that is, $s^{\ast }\neq
1/R_{g}$.

Hence, the next generation matrix can in fact shed light into two properties
of the dynamics: the beginning of the epidemics via the reproduction number
and the approaching to the epidemics level via the asymptotic fraction of
susceptible individuals. The goal of this paper is the description of these
two distinct phases of an epidemics, which is possible if next generation
matrix could be constructed in different ways. In order to show that the
beginning (introduction of infection) and the asymptotic level (final size)
of epidemics are provided by different next generation matrices, drug
sensitive and resistant tuberculosis transmission model is revisited \cite%
{raimundo}. That model was analyzed numerically, which is the reason behind
this revisiting aiming the calculation of both thresholds, which involves a
quite complex manner to evaluate them.

The paper is structured as follows. In section 2, a brief description of the
revisited drug sensitive and resistant tuberculosis model is provided aiming
the calculation of the gross reproduction number and the fraction of
susceptible individuals by different constructions of the next generation
matrix. Discussion is presented in section 3, and Conclusion is given in
section 4.

\section{Material and methods}

The ineffective treatment of tuberculosis leads to emergence of multidrug
resistant (MDR) \textit{Mycobacterium tuberculosis} to the two most potent
first-line medications (isoniazid and rifampin) \cite{frieden}. Tuberculosis
is responsible for the most deaths worldwide, and in 2017, MDR tuberculosis
contributed to $14\%$ of these deaths globally \cite{knight}.

\subsection{Revisiting drug sensitive and resistant tuberculosis model}

In \cite{raimundo} a tuberculosis transmission model was proposed including
drug treatment. They assumed that failure in treatment can arise drug
resistant \textit{M. tuberculosis}, resulting in model%
\begin{equation}
\left\{ 
\begin{array}{rll}
\displaystyle\frac{d}{d\tau }s & = & \mu -\beta _{1}i_{1}s-\beta
_{2}i_{2}s-\mu s \\ 
\displaystyle\frac{d}{d\tau }e_{1} & = & \beta _{1}i_{1}s+\left( 1-q\right)
\xi i_{1}+\eta k_{1}i_{2}-\left( \mu +\gamma \right) e_{1} \\ 
\displaystyle\frac{d}{d\tau }i_{1} & = & \gamma e_{1}+p\gamma e_{2}-\left(
\mu +\alpha +\xi \right) i_{1} \\ 
\displaystyle\frac{d}{d\tau }e_{2} & = & \beta _{2}i_{2}s+\eta
k_{2}i_{2}-\left( \mu +\gamma \right) e_{2} \\ 
\displaystyle\frac{d}{d\tau }i_{2} & = & \left( 1-p\right) \gamma e_{2}+q\xi
i_{1}-\left[ \mu +\alpha +\eta \left( k_{1}+k_{2}\right) \right] i_{2},%
\end{array}%
\right.  \label{tubercul}
\end{equation}%
where the fraction of susceptible individuals is $s$, the fractions of
exposed and infectious with drug sensitive tuberculosis are $e_{1}$ and $%
i_{1}$, and the fractions of exposed and infectious with drug resistant
tuberculosis are $e_{2}$ and $i_{2}$.

Model parameters are briefly described (see \cite{raimundo}). The drug
sensitive and drug resistant transmission rates are $\beta _{1}$ and $\beta
_{2}$. Parameters $\mu $ and $\alpha $ are the natural and tuberculosis
induced mortality rates; $\gamma $ is the endogenous reactivation rate; $\xi 
$ and $\eta $ are drug sensitive and drug resistant treatment rates; $p$ is
the proportion of drug resistant exposed tuberculosis individuals that
develop drug sensitive infectious individuals; $q$ is the probability that
treatment failure occurs due to the development of antibiotic resistance;
and $k_{1}$ and $k_{2}$ are the relative treatment efficacy of drug
sensitive and drug resistant patients.

In \cite{raimundo}, the authors obtained a threshold applying $\mathbf{M}$%
-matrix theory, however, neither gross reproduction number nor the fraction
of susceptible individuals were obtained. Tuberculosis modeling considering
drug sensitive and drug resistant strains presents a little bit complex
calculation of both thresholds.

The system of equations (\ref{tubercul}) has the trivial equilibrium $P^{0}$%
, or disease free equilibrium, given by

\[
P^{0}=\left( \bar{s}=1,\bar{e}_{1}=0,\bar{\imath}_{1}=0,\bar{e}_{2}=0,\bar{%
\imath}_{2}=0\right) ,
\]%
and the non-trivial equilibrium $P^{\ast }$, or endemic equilibrium, given by

\[
P^{\ast }=\left( \bar{s}=s^{\ast },\bar{e}_{1}=e_{1}^{\ast }(s^{\ast }),\bar{%
\imath}_{1}=i_{1}^{\ast }(s^{\ast }),\bar{e}_{2}=e_{2}^{\ast }(s^{\ast }),%
\bar{\imath}_{2}=i_{2}^{\ast }(s^{\ast })\right) , 
\]%
with coordinates (they are written as a function of $s^{\ast }$) being given
by%
\begin{equation}
\left\{ 
\begin{array}{rll}
e_{1}^{\ast }(s^{\ast }) & = & \displaystyle\frac{\left[ \beta _{1}s^{\ast
}+\left( 1-q\right) \xi \right] i_{1}^{\ast }(s^{\ast })+\eta
k_{1}i_{2}^{\ast }(s^{\ast })}{\mu +\gamma } \\ 
i_{1}^{\ast }(s^{\ast }) & = & \displaystyle\frac{\gamma e_{1}^{\ast
}(s^{\ast })+p\gamma e_{2}^{\ast }(s^{\ast })}{\mu +\alpha +\xi } \\ 
e_{2}^{\ast }(s^{\ast }) & = & \displaystyle\frac{\left[ \beta _{2}s^{\ast
}+\eta k_{2}\right] i_{2}^{\ast }(s^{\ast })}{\mu +\gamma } \\ 
i_{2}^{\ast }(s^{\ast }) & = & \displaystyle\frac{\mu +\alpha +\eta -\frac{%
\gamma }{\mu +\gamma }\left[ \beta _{1}s^{\ast }+\left( 1-q\right) \xi %
\right] }{\frac{\gamma }{\mu +\gamma }\left[ \eta k_{1}+p\left( \beta
_{2}s^{\ast }+\eta k_{2}\right) \right] }i_{1}(s^{\ast }),%
\end{array}%
\right.  \label{coordi}
\end{equation}
where the fraction of susceptible individuals $s^{\ast }$ is positive
solution of $Pol(s)=0$, a second degree polynomial given by 
\begin{equation}
\begin{array}{rcl}
Pol(s) & = & R_{10}R_{20}s^{2}-\left[ R_{10}\left( 1-R_{21}\right)
+R_{20}\left( 1-R_{11}\right) +R_{31}\right] s \\ 
&  & \displaystyle+\left( 1-R_{11}\right) \left( 1-R_{21}\right) \left[ 1-%
\frac{R_{32}}{\left( 1-R_{11}\right) \left( 1-R_{21}\right) }\right] ,%
\end{array}
\label{tuber_sus}
\end{equation}%
with the parameters $R_{ij}$ being given by%
\begin{equation}
\left\{ 
\begin{array}{lcccl}
\displaystyle R_{10}=\frac{\gamma }{\mu +\gamma }\frac{\beta _{1}}{\mu
+\alpha +\xi } &  & \mathrm{and} &  & \displaystyle R_{11}=\frac{\xi }{\mu
+\alpha +\xi }\left( 1-q\right) \frac{\gamma }{\mu +\gamma } \\ 
\displaystyle R_{20}=\frac{\gamma }{\mu +\gamma }\left( 1-p\right) \frac{%
\beta _{2}}{\mu +\alpha +\eta \left( k_{1}+k_{2}\right) } &  & \mathrm{and}
&  & \displaystyle R_{21}=\frac{\eta k_{2}}{\mu +\alpha +\eta \left(
k_{1}+k_{2}\right) }\frac{\gamma }{\mu +\gamma }\left( 1-p\right) \\ 
\displaystyle R_{31}=\frac{\gamma }{\mu +\gamma }p\frac{\xi }{\mu +\alpha
+\xi }q\frac{\beta _{2}}{\mu +\alpha +\eta \left( k_{1}+k_{2}\right) } &  & 
\mathrm{and} &  & \displaystyle R_{32}=\frac{\eta \left( k_{1}+pk_{2}\right) 
}{\mu +\alpha +\eta \left( k_{1}+k_{2}\right) }\frac{\gamma }{\mu +\gamma }%
\frac{\xi }{\mu +\alpha +\xi }q,%
\end{array}%
\right.  \label{rij_tuberc}
\end{equation}%
By observing these parameters, it is obvious that $R_{11}<1$, $R_{21}<1$ and 
$R_{32}<1$. The difference $\left( 1-R_{11}\right) \left( 1-R_{21}\right)
-R_{32}$ is written as%
\[
\left( 1-R_{11}\right) \left( 1-R_{21}\right) -R_{32}=\frac{\left( \mu
+\gamma \right) ^{2}\left( \mu +\alpha \right) \left( \mu +\alpha +\eta
k_{1}\right) +d_{1}+d_{2}}{\left( \mu +\gamma \right) ^{2}\left( \mu +\alpha
+\xi \right) \left[ \mu +\alpha +\eta \left( k_{1}+k_{2}\right) \right] }>0, 
\]%
where%
\[
\left\{ 
\begin{array}{l}
d_{1}=\xi \left( \mu +\gamma \right) \left[ \mu \left( \mu +\alpha +\eta
k_{1}\right) +q\gamma \left( \mu +\alpha \right) \right] \\ 
d_{2}=\eta k_{2}\left\{ \left[ \left( \mu +\gamma \right) \left( \mu +\alpha
\right) +\mu \xi \right] \left( \mu +p\gamma \right) +q\left( 1-p\right) \mu
\gamma \xi \right\} ,%
\end{array}%
\right. 
\]%
showing that $R_{32}/\left[ \left( 1-R_{11}\right) \left( 1-R_{21}\right) %
\right] <1$. Notice that $R_{10}$ and $R_{20}$ are the basic reproduction
numbers of drug sensitive and resistant strains of \textit{M. tuberculosis},
and $R_{11}$ and $R_{21}$ are the additional reproduction numbers of drug
sensitive and resistant strains of \textit{M. tuberculosis}. Finally, $%
R_{31} $ and $R_{32}$ are the additional reproduction numbers of resistant
strain of \textit{M. tuberculosis} passing through sensitive strain. (See
Appendix A for interpretation of $R_{ij}$.)

In a transmission model of drug sensitive and resistant \textit{M.
tuberculosis}, $s^{\ast }$ is calculated from the equation $Pol(s)=0$, with $%
Pol(s)$ being given by equation (\ref{tuber_sus}), whose discriminant is%
\[
\begin{array}{ccl}
\Delta  & = & \left[ R_{10}\left( 1-R_{21}\right) +R_{20}\left(
1-R_{11}\right) +R_{31}\right] ^{2}-4R_{10}R_{20}\left[ \left(
1-R_{11}\right) \left( 1-R_{21}\right) -R_{32}\right]  \\ 
& = & \left[ R_{10}\left( 1-R_{11}\right) -R_{20}\left( 1-R_{21}\right)
+R_{31}\right] ^{2}+4R_{20}\left[ R_{10}R_{32}+R_{31}\left( 1-R_{11}\right) %
\right] >0.%
\end{array}%
\]%
Hence, it has always two positive solutions, where the small one is given by%
\begin{equation}
s^{\ast }=s_{s}^{\ast }=\frac{\left[ R_{10}\left( 1-R_{21}\right)
+R_{20}\left( 1-R_{11}\right) +R_{31}\right] -\sqrt{\Delta }}{2R_{10}R_{20}},
\label{susc_geral}
\end{equation}%
which is biologically feasible. The big solution $s_{b}^{\ast }$, given by%
\begin{equation}
s_{b}^{\ast }=\frac{\left[ R_{10}\left( 1-R_{21}\right) +R_{20}\left(
1-R_{11}\right) +R_{31}\right] +\sqrt{\Delta }}{2R_{10}R_{20}},  \label{sbig}
\end{equation}%
does not have biological meaning, which can be noticed from equation (\ref%
{coordi}). The coordinate of individuals with drug sensitive tuberculosis $%
i_{1}^{\ast }$ is always positive, but, rewriting the coordinate of
individuals with drug resistant tuberculosis $i_{2}^{\ast }$ as%
\[
i_{2}^{\ast }=\frac{\left( \mu +\alpha +\xi \right) \left( \mu +\gamma
\right) }{\gamma \left[ \eta \left( k_{1}+pk_{2}\right) +p\beta _{2}s^{\ast }%
\right] }R_{10}\left( s^{m}-s^{\ast }\right) i_{1}^{\ast },
\]%
it is positive whenever $s^{\ast }<s^{m}$, where $s^{m}=\left(
1-R_{11}\right) /R_{10}$. However, evaluating $Pol(s)$ at this value, $%
Pol(s^{m})$ is%
\[
Pol\left( s^{m}\right) =-\left[ \frac{R_{31}}{R_{10}}\left( 1-R_{11}\right)
+R_{32}\right] <0,
\]%
which implies that $s^{m}$ situates between small and big solutions of $%
Pol(s)=0$, or, $s_{s}^{\ast }<s^{m}<s_{b}^{\ast }$, resulting in $%
i_{2}^{\ast }(s_{s}^{\ast })>0$ and $i_{2}^{\ast }(s_{b}^{\ast })<0$. Hence,
all coordinates of $P^{\ast }$ are positive only for small solution $%
s_{s}^{\ast }$, implying that there is a unique non-trivial equilibrium
point.

Let only drug sensitive or resistant strain of \textit{M. tuberculosis}
transmission be considered. This has didactical purpose only (actually, it
does not occur).

Firstly, letting $\beta _{2}=0$ (hence $R_{20}=0$ and $R_{31}=0$),
tuberculosis transmission among individuals is due only by those infected by
drug sensitive strain ($i_{1}^{\ast }$), due to the assumption that
individuals infected by drug resistant strain ($i_{2}^{\ast }$) originated
from failure of drug administration are not transmitting. In this case, the
fraction of susceptible individuals is%
\begin{equation}
s^{\ast }=\frac{1-R_{11}}{R_{10}}\left[ 1-\frac{R_{32}}{\left(
1-R_{11}\right) \left( 1-R_{21}\right) }\right] ,  \label{susc_beta1}
\end{equation}%
showing that the additional decreasing in susceptibles, given by $R_{32}/%
\left[ \left( 1-R_{11}\right) \left( 1-R_{21}\right) \right] $, is due to
the failure of treatment, resulting in non-transmissible (by assumption)
infected individuals with drug resistant strain. If failure in treatment
does not occur, that is, $R_{32}=0$, then the fraction of susceptibles become%
\begin{equation}
s^{\ast }=\frac{1-R_{11}}{R_{10}}=\frac{1}{R_{10}}-\frac{R_{11}}{R_{10}},
\label{senstb}
\end{equation}%
and drug resistant \textit{M. tuberculosis} is not circulating.

Now, letting $\beta _{1}=0$ (hence $R_{10}=0$), tuberculosis transmission
among individuals is due only by those infected by drug resistant strain ($%
i_{2}^{\ast }$), due to the assumption that individuals infected by drug
sensitive strain ($i_{1}^{\ast }$) are not transmitting. In this case, the
fraction of susceptibles is%
\begin{equation}
s^{\ast }=\frac{1-R_{21}}{R_{20}}\left[ 1-\frac{R_{32}}{\left(
1-R_{11}\right) \left( 1-R_{21}\right) }\right] \frac{1}{1+\frac{R_{31}}{%
R_{20}\left( 1-R_{11}\right) }},  \label{susc_beta2}
\end{equation}%
showing again that the additional decreasing in susceptibles, given by $%
R_{32}/\left[ \left( 1-R_{11}\right) \left( 1-R_{21}\right) \right] $, is
due to the failure of treatment, resulting in non-transmissible (by
assumption) infected individuals with drug sensitive strain. In this case,
however, a second additional decreasing in susceptibles appears, given by $%
1/\left\{ 1+R_{31}/\left[ R_{20}\left( 1-R_{11}\right) \right] \right\} $,
due to the passage from $i_{1}$ to $i_{2}$. If failure in treatment and
passage from $i_{1}$ to $i_{2}$ do not occur, that is $R_{31}=R_{32}=0$, the
fraction of susceptibles becomes%
\begin{equation}
s^{\ast }=\frac{1-R_{21}}{R_{20}}=\frac{1}{R_{20}}-\frac{R_{21}}{R_{20}},
\label{resistb}
\end{equation}%
similar to equation (\ref{senstb}), and drug sensitive \textit{M.
tuberculosis} is not circulating.

Notice that when $R_{31}=R_{32}=0$, the dynamics of drug sensitive and
resistant strains of tuberculosis transmissions are decoupled, and each one
can be dealt with separately.

\subsection{Thresholds -- $R_{g}$ and $s^{\ast }$}

In preceding section, the fraction of susceptible individuals at endemic
equilibrium $s^{\ast }$ was evaluated. In this section, this value will be
obtained from the next generation matrix evaluated at the trivial
equilibrium point. Briefly, the next generation matrix is constructed based
on the transmission ($f$) and transition ($v$) vectors, from which matrices,
respectively, $F$ and $V$ evaluated at the trivial equilibrium are obtained,
resulting in the next generation matrix $FV^{-1}$.

In drug sensitive and resistant tuberculosis transmissions model, there are
several next generation matrices. Only two next generation matrices
evaluated at the trivial equilibrium $P^{0}$ are considered, with the
matrices being obtained from the vector of variables $x=\left(
e_{1},i_{1},e_{2},i_{2}\right) ^{T}$, where superscript $T$ stands for the
transposition of a matrix.

\subsubsection{The gross reproduction number $R_{g}$ \label{limiar1}}

In order to obtain the basic reproduction number, diagonal matrix $V$ is
considered. Hence, the vectors $f$ and $v$ are%
\[
\begin{array}{ccccc}
f=\left( 
\begin{array}{c}
\beta _{1}i_{1}s+\left( 1-q\right) \xi i_{1}+\eta k_{1}i_{2} \\ 
\gamma e_{1}+p\gamma e_{2} \\ 
\beta _{2}i_{2}s+\eta k_{2}i_{2} \\ 
\left( 1-p\right) \gamma e_{2}+q\xi i_{1}%
\end{array}%
\right)  &  & \mathrm{and} &  & v=\left( 
\begin{array}{c}
\left( \mu +\gamma \right) e_{1} \\ 
\left( \mu +\alpha +\xi \right) i_{1} \\ 
\left( \mu +\gamma \right) e_{2} \\ 
\left[ \mu +\alpha +\eta \left( k_{1}+k_{2}\right) \right] i_{2}%
\end{array}%
\right) 
\end{array}%
\]%
from which we obtain the matrices $F$ and $V$ given by%
\[
\begin{array}{ccc}
F=\left[ 
\begin{array}{cccc}
0 & \beta _{1}+\left( 1-q\right) \xi  & 0 & \eta k_{1} \\ 
\gamma  & 0 & p\gamma  & 0 \\ 
0 & 0 & 0 & \beta _{2}+\eta k_{2} \\ 
0 & q\xi  & \left( 1-p\right) \gamma  & 0%
\end{array}%
\right]  & \mathrm{and} & V=\left[ 
\begin{array}{cccc}
\mu +\gamma  & 0 & 0 & 0 \\ 
0 & \mu +\alpha +\xi  & 0 & 0 \\ 
0 & 0 & \mu +\gamma  & 0 \\ 
0 & 0 & 0 & \varphi 
\end{array}%
\right] ,%
\end{array}%
\]%
with $\varphi =\mu +\alpha +\eta \left( k_{1}+k_{2}\right) $. The next
generation matrix $FV^{-1}$ is%
\[
FV^{-1}=\left[ 
\begin{array}{cccc}
0 & \displaystyle\frac{\beta _{1}+\left( 1-q\right) \xi }{\mu +\alpha +\xi }
& 0 & \displaystyle\frac{\eta k_{1}}{\mu +\alpha +\eta \left(
k_{1}+k_{2}\right) } \\ 
\frac{\gamma }{\mu +\gamma } & 0 & \displaystyle\frac{p\gamma }{\mu +\gamma }
& 0 \\ 
0 & 0 & 0 & \displaystyle\frac{\beta _{2}+\eta k_{2}}{\mu +\alpha +\eta
\left( k_{1}+k_{2}\right) } \\ 
0 & \displaystyle\frac{q\xi }{\mu +\alpha +\xi } & \displaystyle\frac{\left(
1-p\right) \gamma }{\mu +\gamma } & 0%
\end{array}%
\right] ,
\]%
and the characteristic equation corresponding to $FV^{-1}$ is%
\begin{equation}
\left( \lambda ^{2}-R_{1}\right) \left( \lambda ^{2}-R_{2}\right)
-R_{3}\lambda =0,  \label{charact_gross}
\end{equation}%
where $R_{i}$, with $i=1$, $2$ and $3$, are given by%
\begin{equation}
\left\{ 
\begin{array}{lll}
R_{1} & = & R_{10}+R_{11} \\ 
R_{2} & = & R_{20}+R_{21} \\ 
R_{3} & = & R_{31}+R_{32},%
\end{array}%
\right.   \label{3rs}
\end{equation}%
with $R_{ij}$ being given by equation (\ref{rij_tuberc}). According to \cite%
{yang1}, the gross reproduction number $R_{g}$ is given by%
\begin{equation}
R_{g}=\max \left\{ R_{1},R_{2},\frac{R_{3}}{\left( 1-R_{1}\right) \left(
1-R_{2}\right) }\right\} ,  \label{Rg}
\end{equation}%
where $\max $ stands for the maximum value among them. Notice that the
spectral radius $\rho $ can not be obtained analytically.

The condition to the trivial equilibrium point $P^{0}$ be locally
asymptotically stable (LAS) is $R_{g}<1$. If $R_{g}>1$, $P^{0}$ is unstable,
and the unique non-trivial equilibrium point $P^{\ast }$ appears. Therefore, 
$R_{g}$ is a threshold.

Let two cases be considered. Firstly, consider $R_{32}=0$, that is, there is
not failure in treatment, or $R_{31}=0$, that is, there is not passage from $%
i_{1}$ to $i_{2}$. In this case, $R_{g}$ is%
\[
R_{g}=\max \left\{ R_{1},R_{2},\frac{R_{31}}{\left( 1-R_{1}\right) \left(
1-R_{2}\right) }\mathrm{or}\frac{R_{32}}{\left( 1-R_{1}\right) \left(
1-R_{2}\right) }\right\} , 
\]%
showing that drug sensitive and resistant strains of tuberculosis can reach
endemic level even when $R_{1}<1$ and $R_{2}<1$, but $R_{31}/\left[ \left(
1-R_{1}\right) \left( 1-R_{2}\right) \right] >1\mathrm{\ }$or $R_{32}/\left[
\left( 1-R_{1}\right) \left( 1-R_{2}\right) \right] >1$. The joint
propagation of drug sensitive and resistant strains facilitates the
persistence of epidemics.

However, if $R_{32}=0$ and $R_{31}=0$, $R_{g}$ is%
\[
R_{g}=\max \left\{ R_{1},R_{2}\right\} , 
\]%
both strains propagate independently. Notice that if $R_{1}>1$ and $R_{2}<1$%
, drug sensitive tuberculosis is in endemic level, but drug resistant
tuberculosis goes to extinction, and vice-versa if $R_{1}<1$ and $R_{2}>1$.

\subsubsection{The fraction of susceptible individuals $s^{\ast }$}

In order to obtain the fraction of susceptible individuals, infection matrix 
$M$ must be the simplest (matrix with least number of non-zeros). Hence, the
vectors $f$ and $v$ are%
\[
\begin{array}{ccccc}
f=\left( 
\begin{array}{c}
\beta _{1}i_{1}s \\ 
0 \\ 
\beta _{2}i_{2}s \\ 
0%
\end{array}%
\right) &  & \mathrm{and} &  & v=\left( 
\begin{array}{c}
-\left( 1-q\right) \xi i_{1}-\eta k_{1}i_{2}+\left( \mu +\gamma \right) e_{1}
\\ 
-\gamma e_{1}-p\gamma e_{2}+\left( \mu +\alpha +\xi \right) i_{1} \\ 
-\eta k_{2}i_{2}+\left( \mu +\gamma \right) e_{2} \\ 
-\left( 1-p\right) \gamma e_{2}-q\xi i_{1}+\left[ \mu +\alpha +\eta \left(
k_{1}+k_{2}\right) \right] i_{2}%
\end{array}%
\right)%
\end{array}%
\]%
from which we obtain the matrices $F$ and $V$ given by%
\[
\begin{array}{ccc}
F=\left[ 
\begin{array}{cccc}
0 & \beta _{1} & 0 & 0 \\ 
0 & 0 & 0 & 0 \\ 
0 & 0 & 0 & \beta _{2} \\ 
0 & 0 & 0 & 0%
\end{array}%
\right] & \mathrm{and} & V=\left[ 
\begin{array}{cccc}
\mu +\gamma & -\left( 1-q\right) \xi & 0 & -\eta k_{1} \\ 
-\gamma & \mu +\alpha +\xi & -p\gamma & 0 \\ 
0 & 0 & \mu +\gamma & -\eta k_{2} \\ 
0 & -q\xi & -\left( 1-p\right) \gamma & \mu +\alpha +\eta \left(
k_{1}+k_{2}\right)%
\end{array}%
\right] .%
\end{array}%
\]%
The next generation matrix $FV^{-1}$ is%
\[
FV^{-1}=\left[ 
\begin{array}{cccc}
\beta _{1}n_{11} & \beta _{1}n_{12} & \beta _{1}n_{13} & \beta _{1}n_{14} \\ 
0 & 0 & 0 & 0 \\ 
\beta _{2}n_{31} & \beta _{2}n_{32} & \beta _{2}n_{33} & \beta _{2}n_{34} \\ 
0 & 0 & 0 & 0%
\end{array}%
\right] , 
\]%
($n_{ij}$ are omitted) and the characteristic equation corresponding to $%
FV^{-1}$ is%
\[
\lambda ^{2}\left[ \left( \lambda -\beta _{1}n_{11}\right) \left( \lambda
-\beta _{2}n_{33}\right) -\beta _{1}n_{13}\beta _{2}n_{31}\right] =0, 
\]%
or, letting $\chi _{1}=\beta _{1}n_{11}$, $\chi _{2}=\beta _{2}n_{33}$, $%
\chi _{3}=\beta _{1}n_{13}$, and $\chi _{4}=\beta _{2}n_{31}$,%
\begin{equation}
\lambda ^{2}\left[ \left( \lambda -\chi _{1}\right) \left( \lambda -\chi
_{2}\right) -\chi _{3}\chi _{4}\right] =0,  \label{charac_sus}
\end{equation}%
where $\chi _{i}$ are given by ($p\neq 0$)%
\[
\left\{ 
\begin{array}{lll}
\chi _{1} & = & \displaystyle\frac{R_{10}}{1-R_{11}}\left[ 1-\frac{R_{32}}{%
\left( 1-R_{11}\right) \left( 1-R_{21}\right) }\right] ^{-1} \\ 
\chi _{2} & = & \displaystyle\frac{R_{20}\left( 1-R_{11}\right) +R_{31}}{%
\left( 1-R_{11}\right) \left( 1-R_{21}\right) }\left[ 1-\frac{R_{32}}{\left(
1-R_{11}\right) \left( 1-R_{21}\right) }\right] ^{-1} \\ 
\chi _{3} & = & \displaystyle p\frac{R_{10}}{1-R_{11}}\frac{1+\frac{1}{p}%
R_{33}}{1-R_{21}}\left[ 1-\frac{R_{32}}{\left( 1-R_{11}\right) \left(
1-R_{21}\right) }\right] ^{-1} \\ 
\chi _{4} & = & \displaystyle\frac{1}{p}\frac{R_{20}\left( 1-R_{11}\right)
+R_{31}}{\left( 1-R_{11}\right) \left( 1-R_{21}\right) }\left[ 1-\frac{R_{32}%
}{\left( 1-R_{11}\right) \left( 1-R_{21}\right) }\right] ^{-1},%
\end{array}%
\right. 
\]%
with $R_{ij}$ being given by equation (\ref{rij_tuberc}), and additional $%
R_{33}$ being given by%
\[
R_{33}=\frac{\eta k_{1}}{\mu +\alpha +\eta \left( k_{1}+k_{2}\right) }\frac{%
\gamma }{\mu +\gamma }\left( 1-p\right) . 
\]%
Notice that the parameter $R_{33}$ does not appear in the calculation of the
fraction of susceptible individuals $s^{\ast }$ nor in the gross
reproduction number $R_{g}$.

The characteristic equation (\ref{charac_sus}) has two equal eigenvalues $%
\lambda =0$, and other two are given by solutions of%
\begin{equation}
\lambda ^{2}-\left( \chi _{1}+\chi _{2}\right) \lambda +\chi _{1}\chi
_{2}-\chi _{3}\chi _{4}=0,  \label{charac_sec}
\end{equation}%
which has two positive eigenvalues. (It is easy to show that $\chi _{1}\chi
_{2}-\chi _{3}\chi _{4}>0$ and $\left( \chi _{1}+\chi _{2}\right)
^{2}-4\left( \chi _{1}\chi _{2}-\chi _{3}\chi _{4}\right) >0$.) Hence, the
spectral radius is the big solution, that is,%
\begin{equation}
\rho =\frac{\chi _{1}+\chi _{2}+\sqrt{\left( \chi _{1}+\chi _{2}\right)
^{2}-4\left( \chi _{1}\chi _{2}-\chi _{3}\chi _{4}\right) }}{2}.  \label{rho}
\end{equation}%
Hence, the trivial equilibrium point $P^{0}$ is LAS if $\rho <1$, and $\rho $
is a threshold.

It is clear that this new threshold $\rho $, given by equation (\ref{rho}),
can not be associated with the gross reproduction number $R_{g}$, given by
equation (\ref{Rg}). To clarify the appearance of a second threshold, let,
as the previous section, two special cases be considered.

Firstly, when $\beta _{2}=0$ ($\chi _{2}=\chi _{4}=0$), the spectral radius
of equation (\ref{charac_sus}) is $\rho _{1}$, and equation (\ref{rho})
becomes%
\begin{equation}
\rho _{1}^{-1}=\frac{1-R_{11}}{R_{10}}\left[ 1-\frac{R_{32}}{\left(
1-R_{11}\right) \left( 1-R_{21}\right) }\right] .  \label{rho1}
\end{equation}%
Comparing $\rho _{1}$ with equation (\ref{susc_beta1}), it is clear that $%
\rho _{1}=1/s^{\ast }$. When $\beta _{1}=0$ ($\chi _{1}=\chi _{3}=0$), the
spectral radius of equation (\ref{charac_sus}) is $\rho _{2}$, and equation (%
\ref{rho}) becomes%
\begin{equation}
\rho _{2}^{-1}=\frac{\left( 1-R_{11}\right) \left( 1-R_{21}\right) }{%
R_{20}\left( 1-R_{11}\right) +R_{31}}\left[ 1-\frac{R_{32}}{\left(
1-R_{11}\right) \left( 1-R_{21}\right) }\right] .  \label{rho2}
\end{equation}%
Comparing $\rho _{2}$ with equation (\ref{susc_beta2}), it is clear that $%
\rho _{2}=1/s^{\ast }$.

It is not an easy task to prove that $\rho =1/s^{\ast }$, when $\beta _{1}>0$
and $\beta _{2}>0$, with $\rho $ and $s^{\ast }$ being given by equations,
respectively, (\ref{rho}) and (\ref{susc_geral}). The main reason is the
parameter $R_{33}$ appearing in $\rho $ but not in $s^{\ast }$, but
numerically $\rho =1/s^{\ast }$ can be verified. Hence, the spectral radius
is exactly the inverse of the fraction of susceptible individuals.

The condition to the trivial equilibrium point $P^{0}$ be LAS is $\rho <1$.
If $\rho >1$, $P^{0}$ is unstable, and the unique non-trivial equilibrium
point $P^{\ast }$ appears. Hence, $P^{\ast }$ is biologically feasible if $%
\rho >1$, that is, $s^{\ast }<1$, and $1/s^{\ast }$ is another threshold.

Finally, letting $R_{32}=0$ besides $\beta _{2}=0$, the spectral radius (\ref%
{rho1}) becomes%
\begin{equation}
\rho _{1}^{-1}=\frac{1-R_{11}}{R_{10}},  \label{rho1a}
\end{equation}%
which is equal to the fraction of susceptibles given by equation (\ref%
{senstb}). For this reason, $R_{10}$ is the basic reproduction number of
drug sensitive strain, and $R_{11}$ is the additional reproduction number.
Now, letting $R_{32}=0$ and $R_{31}=0$ besides $\beta _{1}=0$, the spectral
radius (\ref{rho2}) becomes%
\[
\rho _{2}^{-1}=\frac{1-R_{21}}{R_{20}}, 
\]%
which is equal to the fraction of susceptibles given by equation (\ref%
{resistb}), and $R_{20}$ is the basic reproduction number of drug resistant
strain, and $R_{21}$ is the additional reproduction number.

\section{Discussion}

Tuberculosis modeling considering sensitive and drug resistant \textit{M.
tuberculosis} transmissions was taken as an example of application of the
next generation matrix to describe both the beginning and ending phases of
epidemics.

Depending on the construction of vectors $f$ and $v$, consequently matrices $%
M$ and $V$, two thresholds are obtained from the characteristic equations
corresponding to the next generation matrix $FV^{-1}$. Traditionally, the
spectral radius was taken as the basic (gross) reproduction number \cite%
{driessche} \cite{shuai}, but different approach was proposed in \cite{yang1}%
, which consists in summing the coefficients of the characteristic equation
rather than evaluating the spectral radius. This approach has a substantial
advantage: there is not necessity of a recipe to construct vectors $f$ and $v
$ \cite{roberts}.

However, depending on the complexity of the model, the sum of the
coefficients is not sufficient to determine the gross reproduction number.
The model of tuberculosis revisited here is an example. The method used to
obtain two thresholds $R_{g}$ and $1/s^{\ast }$ is summarized: Let the
characteristic equation corresponding to next generation matrix $FV^{-1}$ be
written as%
\begin{equation}
\Lambda (\lambda )=\Lambda _{n}(\lambda )\Lambda _{m}(\lambda )-\Lambda
_{p}(\lambda ),  \label{lambda_more}
\end{equation}%
where $\Lambda _{n}(\lambda )=\Lambda _{n}(\lambda )=\lambda
^{n}-a_{n-1}\lambda ^{n-1}-\cdots -a_{1}\lambda -a_{0}$, $\Lambda
_{m}(\lambda )=\lambda ^{m}-b_{m-1}\lambda ^{m-1}-\cdots -b_{1}\lambda
-b_{0} $, and $\Lambda _{p}(\lambda )=$ $c_{p}\lambda ^{p}+\cdots
+c_{1}\lambda +c_{0}$, with $\Omega _{n}=\sum_{i=0}^{n-1}a_{i}$, $\Omega
_{m}=\sum_{i=0}^{m-1}b_{i}$ and $\Omega _{p}=\sum_{i=0}^{p}c_{i}$ (all
coefficients are non-negative).

\begin{description}
\item[(A)] If vector $f$ carries only bilinear terms regarding infection,
and all terms are left to vector $v$ (matrix $F$ has the least number of
non-zeros, while matrix $V$ has the most number of non-zeros), then the
spectral radius $\rho =\rho \left( FV^{-1}\right) $ of the characteristic
equation $\Lambda (\lambda )=0$ is the inverse of the fraction of
susceptible individuals $s^{\ast }$, that is, $s^{\ast }=1/\rho $.

\item[(B)] In all other constructions of vectors $f$ and $v$, the sum of the
coefficients of the characteristic equation corresponding to the next
generation matrix $FV^{-1}$ is the gross reproduction number $R_{g}$, that
is,%
\begin{equation}
R_{g}=\max \left\{ \Omega _{n},\Omega _{m},\frac{\Omega _{p}}{\left(
1-\Omega _{n}\right) \left( 1-\Omega _{m}\right) }\right\} ,  \label{basic}
\end{equation}%
where $\max $ stands for the maximum value among $\Omega _{n}$, $\Omega _{m}$
and $\Omega _{p}/\left[ \left( 1-\Omega _{n}\right) \left( 1-\Omega
_{m}\right) \right] $. Hence, the best choice of construction of vectors $f$
and $v$ is such that the matrix $V$ must be diagonal.
\end{description}

Observe that equation (\ref{lambda_more}) has at least two positive
solutions (excluding the possibility of absence of positive solution). For
this reason, the threshold in item (A) must be the spectral radius. However,
if there are not interactions between pathogens or strains, that is, $%
\Lambda _{p}(\lambda )=0$, then there is a unique positive solution for each
equation $\Lambda _{n}(\lambda )=0$ or $\Lambda _{m}(\lambda )=0$, and $%
s^{\ast }=1/\Omega _{n}$ or $s^{\ast }=1/\Omega _{m}$, instead of spectral
radius. This is called simplified item (A).

Notice that items (A) and (B) were cited in \cite{yang1} but only item (B)
was briefly exemplified (section \ref{limiar1} is direct application of this
item). Here, more details regarding the application of item (B) in
tuberculosis modeling encompassing drug sensitive and resistant strains were
presented, and item (A) is a novel application.

Item (A) dealing with the fraction of susceptible individuals deserves some
comments. The steady state fraction of susceptible individuals is obtained
as the roots of the second degree polynomial (\ref{susc_geral}), which has
two positive solutions. It was shown that only the small one was chosen due
to biological meaning (the big solution generates negative coordinates for
the non-trivial equilibrium). The stability of the trivial equilibrium point
is assessed also by roots of a second degree polynomial, given by the
characteristic equation (\ref{charac_sec}) presenting two positive
solutions. Hence, two reasons are behind the relationship between the
spectral radius and the fraction of susceptible individuals.

\begin{enumerate}
\item When a characteristic equation has more than one positive eigenvalue,
the spectral radius $\rho $ must be chosen as the threshold.

\item The trivial equilibrium is locally asymptotically stable if the
spectral radius is lower than one ($\rho <1$), and unstable otherwise.
Hence, epidemics is settle at the community if $\rho >1$, that is, $s^{\ast
}<1$.
\end{enumerate}

Hence, in epidemics situation, the spectral radius guarantees value higher
than one, and, consequently, the inverse is lower than one. It is not an
easy task to demonstrating analytically that the small solution of equation (%
\ref{susc_geral}) is equal to the inverse of the spectral radius of
characteristic equation (\ref{charac_sec}), $s_{s}^{\ast }=1/\rho $, but it
can be verified numerically. However, in special cases, this relationship
was demonstrated analytically.

For instance, letting $\mu =0.0154$, $\alpha =0.33$, $\gamma =0.025$, $\xi
=0.1$, $\eta =0.5$, $\beta _{1}=4,55$ and $\beta _{2}=6.25$ (all in $%
years^{-1}$); and $p=0.05$, $q=0.4$, $k_{1}=0.87$ and $k_{2}=0.53$
(dimensionless), the reproduction numbers are, from equation (\ref{3rs}), $%
R_{1}=6.4$, $R_{2}=3.7$ and $R_{3}=0.04$, and $R_{in}=0.003$, where $%
R_{in}=R_{32}/\left[ \left( 1-R_{1}\right) \left( 1-R_{2}\right) \right] $.
The small and big fraction of susceptible individuals are, from(\ref%
{susc_geral}) and (\ref{sbig}), $s_{s}^{\ast }=0.1341$ and $s_{g}^{\ast
}=0.2537$. From equation (\ref{rho}), the inverse of spectral radius is $%
1/\rho =0.1314$, while the inverse of small eigenvalue of equation (\ref%
{charac_sec}) is $0.2537$. Hence, the inverse of the spectral radius is
equal to the small fraction of susceptible individuals, which value is in
accordance with asymptotic value obtained by Runge-Kutta method.

It is worth stressing the fact that characteristic equations (\ref%
{charact_gross}) and (\ref{charac_sus}) have similar structure. However, the
gross reproduction number is given by the sum of the coefficients of
equation (\ref{Rg}), while the inverse of the fraction of susceptible
individuals is given by the spectral radius of equation (\ref{charac_sec}).

The special case letting $\beta _{2}=0$ and $R_{32}=0$ dealt with in
preceding section is quite similar to that model considered by Driessche and
Watmough \cite{driessche}. In their analysis they did not realize the
existence of two thresholds, for this reason they considered that the basic
reproduction number is given by equation (\ref{rho1a}), not equation (\ref%
{3rs}).

Is the spectral radius indeed the inverse of the fraction of susceptible
individuals? Let two examples be considered, SEIR and dengue with
transovarial transmission. In both examples, there is only one pathogen,
hence $\Lambda _{p}(\lambda )=0$ in equation (\ref{lambda_more}), resulting
in $\Lambda (\lambda )=\Lambda _{n}(\lambda )$, with $\Omega
_{n}=\sum_{i=0}^{n-1}a_{i}$ and $\Lambda _{n}(\lambda )=0$ has only one
positive solution. For this reason, first threshold is $R_{g}=\Omega _{n}$,
from equation (\ref{basic}), and second threshold is $s^{\ast }=1/\Omega
_{n} $, not the spectral radius. Hence, simplified item (A) must be applied.

Firstly, let the well known SEIR model be considered (see for instance \cite%
{anderson}). The model describes a pathogen being transmitted directly from
infectious to susceptible individuals, which is given by%
\begin{equation}
\left\{ 
\begin{array}{rll}
\displaystyle\frac{d}{dt}s & = & \mu -\beta si-\mu s \\ 
\displaystyle\frac{d}{dt}e & = & \beta si-\left( \mu +\gamma \right) e \\ 
\displaystyle\frac{d}{dt}i & = & \gamma e-\left( \mu +\sigma \right) i \\ 
\displaystyle\frac{d}{dt}r & = & \sigma i-\mu r,%
\end{array}%
\right.  \label{seir}
\end{equation}%
where $s$, $e$, $i$ and $r$ are the fractions of, respectively, susceptible,
exposed, infectious and recovered individuals. The model parameters are the
mortality rate $\mu $, the contact rate $\beta $, the infectious ($\gamma $)
and recovery ($\sigma $) rates.

The system of equations (\ref{seir}) has two equilibrium points: the trivial 
$P^{0}=\left( 1,0,0,0\right) $ and the non-trivial $P^{\ast }=\left( s^{\ast
},e^{\ast },i^{\ast },r^{\ast }\right) $, where $s^{\ast }=1/R_{0}$, with
the basic reproduction number $R_{0}$ being given by%
\begin{equation}
R_{0}=\frac{\gamma }{\mu +\gamma }\times \frac{\beta }{\mu +\sigma }.
\label{r0seir}
\end{equation}

The next generation matrix is obtained considering the vector of variables $%
x=\left( e,i\right) ^{T}$, where superscript $T$ stands for the
transposition of a matrix. In this model, there are only two next generation
matrices evaluated at the trivial equilibrium $P^{0}$.

The basic reproduction number is obtained according to item (B), that is,
the sum of the coefficients of the characteristic equation. The
characteristic equation corresponding to the next generation matrix obtained
from diagonal matrix $V$ is%
\begin{equation}
\lambda ^{2}-R_{0}=0,  \label{charac1}
\end{equation}%
where the basic reproduction number $R_{0}$ is given by equation (\ref%
{r0seir}).

Let procedure stated in simplified item (A) be applied. The characteristic
equation corresponding to the next generation matrix obtained from
non-diagonal matrix $V$ is%
\begin{equation}
\lambda \left( \lambda -R_{0}\right) =0,  \label{charac2}
\end{equation}%
where $R_{0}$ is given by equation (\ref{r0seir}). According to simplified
item (A), this full matrix $V$ must originate the second threshold $%
1/s^{\ast }$ as the sum of coefficients. Hence, the inverse of $R_{0}$ is
the fraction of susceptible individuals, that is, $s^{\ast }=1/R_{0}$ which
appears implicitly. Notice in SEIR model the spectral radius is also $R_{0}$%
, that is, $\rho =R_{0}$.

Notice that in SEIR model, the spectral radius $\rho $ of equation (\ref%
{charac1}) is $\rho =\sqrt{R_{0}}$, while the spectral radius of equation (%
\ref{charac2}) is $\rho =R_{0}$. Notice that $\rho =R_{0}$ is the reason why
some authors claim that the construction of vectors $f$ and $v$ according to
simplified item (A) is correct \cite{driessche}. But, the sum of
coefficients of both equations is the same, that is, $\Omega _{2}=R_{0}$. In
SIR\ model, however, there is only one characteristic equation (the next
generation matrix is a unitary matrix), and the spectral radius is indeed
the basic reproduction number, and there is not a second threshold. Hence,
the fraction of susceptible individuals being the inverse of the basic
reproduction number appears only from the equilibrium value of $s^{\ast }$.

A second model is dengue encompassing transovarial transmission model \cite%
{yang}, which is also revisited to illustrate the existence of two
thresholds describing two extremes: beginning and ending of epidemics. In
SEIR model, there is one pathogen, one population and one route of
transmission, while in dengue with transovarial transmission model, there
are two populations, one common pathogen, but two routes of transmission.

The model presented in \cite{yang} considered dengue virus being transmitted
by both horizontal and transovarial transmission routes. That model was
described by the system of differential equations%
\begin{equation}
\left\{ 
\begin{array}{rll}
\displaystyle\frac{d}{dt}l_{1} & = & qf\phi \left[ m_{1}+\left( 1-\alpha
\right) m_{2}\right] \left( 1-\frac{l_{1}+l_{2}}{C}\right) -\left( \sigma
_{a}+\mu _{a}\right) l_{1} \\ 
\displaystyle\frac{d}{dt}l_{2} & = & qf\phi \alpha m_{2}\left( 1-\frac{%
l_{1}+l_{2}}{C}\right) -\left( \sigma _{a}+\mu _{a}\right) l_{2} \\ 
\displaystyle\frac{d}{dt}m_{1} & = & \sigma _{a}l_{1}-\left( \beta _{m}\phi
i+\mu _{f}\right) m_{1} \\ 
\displaystyle\frac{d}{dt}m_{2} & = & \sigma _{a}l_{2}+\beta _{m}\phi
im_{1}-\mu _{f}m_{2} \\ 
\displaystyle\frac{d}{dt}s & = & \mu _{h}-\left( \frac{\beta _{h}\phi }{N}%
m_{2}+\mu _{h}\right) s \\ 
\displaystyle\frac{d}{dt}i & = & \frac{\beta _{h}\phi }{N}m_{2}s-\left(
\sigma _{h}+\mu _{h}\right) i,%
\end{array}%
\right.  \label{system_to}
\end{equation}%
where the decoupled fraction of immune humans is given by $r=1-s-i$, $s$ and 
$i$ are the fractions of susceptible and infectious humans, and $N$ is the
constant total number of the humans. The susceptible and infectious female
adult mosquitoes are $m_{1}$ and $m_{2}$, with $m=m_{1}+m_{2}$, and $l_{1}$
and $l_{2}$ represent the uninfected and infected immatures, with $%
l=l_{1}+l_{2}$.

With respect to the model parameters, $\alpha $ is the proportion of
transovarial transmission, $\mu _{h}$ is the birth and mortality rates of
humans, and $\sigma _{h}$ is recovery rate. The per-capita oviposition rate
is $\phi $, $q$ and $f$ are the fractions of eggs that are hatching to larva
and that will originate female mosquitoes, respectively, $C$ is the carrying
capacity of the breeding sites, $\sigma _{a}$ is rate at which larva become
adults, and $\mu _{a}$ and $\mu _{f}$ are the mortality rates of,
respectively, immatures and adults. Finally, $\beta _{h}$ is the
transmission coefficient from mosquito to human, and $\beta _{m}$ is the
transmission coefficient from human to mosquito.

The system of equations (\ref{system_to}) has two equilibrium points,
assuming that $Q_{0}>1$, where $Q_{0}=\sigma _{a}qf\phi /\left[ \left(
\sigma _{a}+\mu _{a}\right) \mu _{f}\right] $ is the basic offspring number.
The trivial equilibrium $P^{0}$, or disease free equilibrium, is given by

\begin{equation}
P^{0}=\left( \bar{l}_{1}=l^{\ast }=C\left( 1-\frac{1}{Q_{0}}\right) ,\bar{l}%
_{2}=0,\bar{m}_{1}=m^{\ast }=\frac{\sigma _{a}}{\mu _{f}}C\left( 1-\frac{1}{%
Q_{0}}\right) ,\bar{m}_{2}=0,\bar{s}=1,\bar{\imath}=0\right) ,
\label{mosq_triv_to}
\end{equation}%
and the non-trivial equilibrium $P^{\ast }$, or endemic equilibrium, is
given by

\[
P^{\ast }=\left( \bar{l}_{1}=l_{1}^{\ast },\bar{l}_{2}=l_{2}^{\ast },\bar{m}%
_{1}=m_{1}^{\ast },\bar{m}_{2}=m_{2}^{\ast },\bar{s}=s^{\ast },\bar{\imath}%
=i^{\ast }\right) ,
\]%
where the the product of the fractions of susceptible humans $s^{\ast }$ and
mosquitoes $m_{1}^{\ast }/m^{\ast }$ is%
\begin{equation}
s^{\ast }\times \frac{m_{1}^{\ast }}{m^{\ast }}=\frac{1-R_{v}}{R_{0}}=\frac{1%
}{R_{0}}-\frac{\alpha }{R_{0}}.  \label{equil_fractions_to}
\end{equation}%
(see \cite{yang} for detailed calculations.) The gross reproduction number $%
R_{g}$ is defined as%
\begin{equation}
R_{g}=R_{0}+R_{v},  \label{R_ovarian}
\end{equation}%
which is the sum of the basic reproduction number $R_{0}=R_{0}^{h}R_{0}^{m}$
due to the horizontal transmission with two partial contributions $%
R_{0}^{h}=\beta _{h}\phi /\mu _{f}$ and $R_{0}^{m}=\beta _{m}\phi m^{\ast }/%
\left[ \left( \sigma _{h}+\mu _{h}\right) N\right] $, and the additional
reproduction number $R_{v}=\alpha $ due to the transovarial transmission.

Only two next generation matrices evaluated at the trivial equilibrium $%
P^{0} $ are considered, with the matrices being obtained from the vector of
variables $x=\left( m_{2},i,l_{2}\right) ^{T}$, where superscript $T$ stands
for the transposition of a matrix.

In order to obtain the gross reproduction number, diagonal matrix $V$ is
considered, according to item (B). The next generation matrix $F_{1}V^{-1}$
is%
\[
FV^{-1}=\left[ 
\begin{array}{ccc}
0 & NR_{0}^{m} & \frac{\sigma _{a}}{\sigma _{a}+\mu _{a}} \\ 
\frac{1}{N}R_{0}^{h} & 0 & 0 \\ 
\alpha \frac{\sigma _{a}+\mu _{a}}{\sigma _{a}} & 0 & 0%
\end{array}%
\right] ,
\]%
and the corresponding characteristic equation is%
\begin{equation}
\lambda \left( \lambda ^{2}-R_{g}\right) =0,  \label{eigen2}
\end{equation}%
with $R_{g}$ being given by equation (\ref{R_ovarian}), which is the gross
reproduction number (the sum of the coefficients of the characteristic
equation). In transovarial dengue transmission model, there are other next
generation matrices resulting in the same gross reproduction number (see 
\cite{yang}).

In order to obtain the fraction of susceptible individuals, infection matrix 
$M$ must be the simplest (matrix with least number of non-zeros), thus
matrix $V$ is the most full with non-zero elements. In this case, the next
generation matrix $FV^{-1}$ is%
\[
FV^{-1}=\left[ 
\begin{array}{ccc}
0 & NR_{0}^{m} & 0 \\ 
\displaystyle\frac{1}{1-\alpha }\frac{1}{N}R_{0}^{h} & 0 & \displaystyle%
\frac{1}{1-\alpha }\frac{\sigma _{a}}{\sigma _{a}+\mu _{a}}\frac{1}{N}%
R_{0}^{h} \\ 
0 & 0 & 0%
\end{array}%
\right] . 
\]%
and the characteristic equation corresponding to $FV^{-1}$ is%
\begin{equation}
\lambda \left( \lambda ^{2}-\frac{R_{0}}{1-R_{v}}\right) =0.  \label{eigen1}
\end{equation}%
According to simplified item (A), the sum of the coefficients is a
threshold, that is, $1/\Omega _{3}=(1-R_{v})/R_{0}$. Comparing, however,
with equation (\ref{equil_fractions_to}), the product of the fractions of
susceptible humans and mosquitoes is indeed the threshold, that is, $s^{\ast
}\times m_{1}^{\ast }/m^{\ast }=(1-R_{v})/R_{0}$. This threshold must be the
product of susceptible populations, due to the fact that two populations are
involved in the transmission.

Comparing equations (\ref{charac2}) and (\ref{charac2}) in SEIR model, the
relationship $s^{\ast }=1/R_{0}$ is obeyed, while, from equations (\ref%
{eigen2}) and (\ref{eigen1}), the product $s^{\ast }\times m_{1}^{\ast
}/m^{\ast }$ is not inverse of $R_{g}$. Hence, two routes of transmission
result in two different thresholds. However, if only one route of
transmission is considered, letting $\alpha =0$, then $s^{\ast }\times
m_{1}^{\ast }/m^{\ast }=1/R_{0}$, implying that there is a unique threshold $%
R_{0}$.

\section{Conclusion}

The basic reproduction number has a well accepted interpretation: The
secondary cases produced by one infectious individual when introduced in a
completely susceptible population. This concept portraits the beginning of
an epidemics. Nevertheless, if the next generation matrix provides the
initial strength of an epidemics, it is expected that it may also predict
the final size of an epidemics, which is indeed measured by the remaining
fraction of susceptible individuals -- this fraction portraits the ending
phase of an epidemics, that is, those individuals who have not been infected
at steady state. For instance, if there is only one threshold, the basic
reproduction number $R_{0}$ and the final size of epidemics $s^{\ast }$ obey 
$s^{\ast }=1/R_{0}$. In another words, how intense is an epidemics (higher $%
R_{0}$) more individuals are infected and low number of individuals are left
uninfected, hence the fraction os susceptible individuals is low ($1/R_{0}$).

The procedures presented in \cite{yang1} can be easily applied when the
characteristic equation corresponding to the next generation matrix is given
by equation (\ref{lambda_more}), with $\Lambda _{p}(\lambda )=0$. In this
case, the sum of the coefficients of this equation is the basic (gross)
reproduction number or the fraction of susceptible individuals, as SEIR and
dengue with transovarial models showed. However, when the characteristic
equation corresponding to the next generation matrix is given by equation (%
\ref{lambda_more}), then the gross reproduction number is given by equation (%
\ref{basic}), and the spectral radius is the inverse of the fraction of
susceptible individuals. This case was shown revisiting drug sensitive and
resistant tuberculosis transmission model.

It is worth stressing the fact that the sum of the coefficients of the
characteristic equation of the next generation matrix provides only one
threshold, by the fact that this equation has a unique positive eigenvalue.
When there is not a unique positive eigenvalue, it is natural choosing the
spectral radius for two reasons: (1) it is the greatest value assuming value
higher than one to maintain epidemics, and, consequently, (2) the inverse of
this number is the lowest, which is lower than one. In the case of the
fraction of susceptible individuals, this number must be lower than one.

It is well accepted the fact that the basic (gross) reproduction number
obtained from the next generation matrix is linked to the initial phase of
an epidemics. Also, the global stability of the trivial equilibrium point
could be determined by the left eigenvector associated to this next
generation matrix \cite{shuai}. Besides these two important results, the
next generation matrix can predict the final size of an epidemics by
allowing the calculation of the steady state fraction of susceptible
individuals. Therefore, depending on how the next generation matrix is
constructed, both initial and final phases of an epidemics can be estimated.



\appendix

\renewcommand{\theequation}{\Alph{section}.\arabic{equation}} %
\setcounter{equation}{0}

\section{Interpreting $R_{ij}$}

In \cite{raimundo} neither the gross reproduction number nor the fraction of
susceptible individuals were obtained. Hence, the interpretations of $R_{ij}$
given by equation (\ref{rij_tuberc}) are done.

\begin{description}
\item[A.] Drug sensitive tuberculosis transmission $R_{1}=R_{10}+R_{11}$.
\end{description}

\begin{enumerate}
\item $\displaystyle R_{10}=\frac{\gamma }{\mu +\gamma }\times \frac{\beta
_{1}}{\mu +\alpha +\xi }$. A primary drug sensitive infectious individual
survives the exposed class $e_{1}$ ($\gamma /\left( \mu +\gamma \right) $),
and during the infectious period in $i_{1}$ generates drug sensitive
secondary cases ($\beta _{1}/\left( \mu +\alpha +\xi \right) $).

\item $\displaystyle R_{11}=\frac{\xi }{\mu +\alpha +\xi }\left( 1-q\right) 
\frac{\gamma }{\mu +\gamma }$. A secondary drug sensitive infectious
individual survives the infectious class $i_{1}$ ($\xi /\left( \mu +\alpha
+\xi \right) $), and a fraction $1-q$ goes back to exposed class $e_{1}$,
surviving this class ($\gamma /\left( \mu +\gamma \right) $) returns to
infectious class $i_{1}$ and generates new cases of sensitive tuberculosis.
\end{enumerate}

\begin{description}
\item[B.] Drug resistant tuberculosis transmission $R_{2}=R_{20}+R_{21}$.
\end{description}

\begin{enumerate}
\item $\displaystyle R_{20}=\frac{\gamma }{\mu +\gamma }\left( 1-p\right) 
\frac{\beta _{2}}{\mu +\alpha +\eta \left( k_{1}+k_{2}\right) }$. A primary
drug resistant infectious individual survives the exposed class $e_{2}$ ($%
\gamma /\left( \mu +\gamma \right) $), and a proportion $1-p$ enters to $%
i_{2}$, and during the infectious period generates drug resistant secondary
cases ($\beta _{2}/\left[ \mu +\alpha +\eta \left( k_{1}+k_{2}\right) \right]
$).

\item $\displaystyle R_{21}=\frac{\eta k_{2}}{\mu +\alpha +\eta \left(
k_{1}+k_{2}\right) }\frac{\gamma }{\mu +\gamma }\left( 1-p\right) $. A
secondary drug resistant infectious individual survives the infectious class 
$i_{2}$ ($\eta k_{2}/\left[ \mu +\alpha +\eta \left( k_{1}+k_{2}\right) %
\right] $), goes back to exposed class $e_{2}$ and survives this class ($%
\gamma /\left( \mu +\gamma \right) $), and a fraction $1-p$ returns to
infectious class $i_{2}$ and generates new cases of resistant tuberculosis.
\end{enumerate}

\begin{description}
\item[C.] Drug resistant tuberculosis transmission through drug sensitive
transmission $R_{3}=R_{31}+R_{32}$.
\end{description}

\begin{enumerate}
\item $\displaystyle R_{31}=\frac{\gamma }{\mu +\gamma }p\frac{\xi }{\mu
+\alpha +\xi }q\frac{\beta _{2}}{\mu +\alpha +\eta \left( k_{1}+k_{2}\right) 
}$. A primary drug resistant infectious individual survives the exposed
class $e_{2}$ ($\gamma /\left( \mu +\gamma \right) $), a proportion $p$
enters to $i_{1}$, surviving this class ($\xi /\left( \mu +\alpha +\xi
\right) $) a fraction $q$ goes direct to infectious class $i_{2}$,\ and
during the infectious period generates drug resistant secondary cases ($%
\beta _{2}/\left[ \mu +\alpha +\eta \left( k_{1}+k_{2}\right) \right] $).

\item $\displaystyle R_{32}=\frac{\eta \left( k_{1}+pk_{2}\right) k_{2}}{\mu
+\alpha +\eta \left( k_{1}+k_{2}\right) }\frac{\gamma }{\mu +\gamma }\frac{%
\xi }{\mu +\alpha +\xi }q$. This is split in $R_{321}$ and $R_{322}$.
\end{enumerate}

\begin{description}
\item[2.1.] $\displaystyle R_{321}=\frac{\eta k_{1}}{\mu +\alpha +\eta
\left( k_{1}+k_{2}\right) }\frac{\gamma }{\mu +\gamma }\frac{\xi }{\mu
+\alpha +\xi }q$. A secondary drug resistant infectious individual survives
the infectious class $i_{2}$ ($\eta k_{1}/\left[ \mu +\alpha +\eta \left(
k_{1}+k_{2}\right) \right] $), goes back to exposed class $e_{1}$ and
survives this class ($\gamma /\left( \mu +\gamma \right) $), and enters to
infectious class $i_{1}$ and surviving this class ($\xi /\left( \mu +\alpha
+\xi \right) $) a fraction $q$ returns to infectious class $i_{2}$ and
generates resistant tuberculosis.

\item[2.2.] $\displaystyle R_{322}=\frac{\eta k_{2}}{\mu +\alpha +\eta
\left( k_{1}+k_{2}\right) }\frac{\gamma }{\mu +\gamma }p\frac{\xi }{\mu
+\alpha +\xi }q$. A secondary drug resistant infectious individual survives
the infectious class $i_{2}$ ($\eta k_{2}/\left[ \mu +\alpha +\eta \left(
k_{1}+k_{2}\right) \right] $), goes back to exposed class $e_{2}$ and
survives this class ($\gamma /\left( \mu +\gamma \right) $), and a fraction $%
p$ enters to infectious class $i_{1}$, surviving this class ($\xi /\left(
\mu +\alpha +\xi \right) $) a fraction $q$ returns to infectious class $%
i_{2} $ and generates resistant tuberculosis.
\end{description}

\end{document}